\documentclass[prl,nofootinbib,longbibliography,twocolumn]{revtex4-2}
\usepackage{amsmath}
\usepackage{amssymb}
\usepackage{graphicx}
\usepackage{dcolumn}
\usepackage{bm}
\usepackage[colorlinks=true,linkcolor=blue,anchorcolor=blue, citecolor=cyan,urlcolor=cyan]{hyperref}
\usepackage[mathlines]{lineno}
\usepackage{ulem}
\usepackage{pifont}

\newcommand{\bfk}{\boldsymbol{k}}

\newcommand{\cs}[1]{\left(#1\right)}

\newcommand{\bsf}[1]{\boldsymbol{#1}}
\begin{document}
    \title{Tunable Topological Superconductivity by Fully Compensated Ferrimagnets}
    \author{Yu-Xuan Li}
    \author{Yicheng Liu}
    \author{Cheng-Cheng Liu}
    \email{ccliu@bit.edu.cn}
    \affiliation{Centre for Quantum Physics, Key Laboratory of Advanced Optoelectronic Quantum Architecture and Measurement (MOE), School of Physics, Beijing Institute of Technology, Beijing 100081, China}
\begin{abstract}
We propose a platform based on a fully compensated ferrimagnet (fFIM) for realizing and controlling topological superconductivity with Majorana bound states across multiple dimensions. Through symmetry analysis and microscopic modeling, we demonstrate that fFIM-based heterostructures host (i) Majorana zero modes localized at the ends of one-dimensional nanowires, (ii) chiral Majorana edge states along two-dimensional boundaries, and (iii) tunable Majorana corner modes in higher-order topological phases. The unique properties of fFIMs enable an electric field to drive topological superconductivity phase transitions and N\'eel vector orientation to control the spatial distribution of Majorana modes, without external magnetic fields. Crucially, the absence of net magnetization in fFIM-based heterostructures preserves superconductivity, circumventing the usual trade-off between tunability and superconducting coherence in magnetized systems. Our results establish fFIM-based heterostructures as a versatile platform for tunable topological superconductivity.
\end{abstract}
\maketitle
\textit{Introduction.---}
The interplay between superconductivity and magnetism has emerged as a key mechanism for stabilizing Majorana bound states (MBSs), thereby enabling robust platforms for non-Abelian quantum operations in topological systems~\cite{hasan_colloquium_2010,qi_topological_2011,alicea_new_2012,stanescu_majorana_2013,sato_topological_2017,fu_superconducting_2008,alicea_majorana_2010,linder_unconventional_2010,sau_generic_2010,oreg_helical_2010,lutchyn_majorana_2010,tewari_topological_2012,nakosai_topological_2012,qi_chiral_2010,alicea_non-abelian_2011,potter_majorana_2011,liu_d_2013,nadj-perge_observation_2014,jeon_distinguishing_2017,pientka_topological_2017,savary_superconductivity_2017,fornieri_evidence_2019,ning_flexible_2024}. While intrinsic topological superconductors (TSCs) remain elusive, engineered heterostructures—particularly one-dimensional (1D) nanowires and two-dimensional (2D) Rashba systems—have provided a promising path toward realizing effective TSCs~\cite{sau_generic_2010,oreg_helical_2010,lutchyn_majorana_2010,nakosai_topological_2012,tewari_topological_2012}. Recent developments in higher-order topology have expanded the theoretical landscape, offering novel routes to Majorana zero modes (MZMs) localized at boundaries with reduced dimensionality, such as corners or hinges~\cite{zhang_surface_2013,langbehn_reflection-symmetric_2017,benalcazar_electric_2017,song__2017,khalaf_higher-order_2018,zhu_topological_2023,ghorashi_altermagnetic_2024,pan_majorana_2024}. However,  most realizations fundamentally rely on magnetization  or  magnetic fields to induced the required spin splitting. The associated stray magnetic fields and depairing inevitably weaken superconductivity and destabilize the topological gap~\cite{buzdin_proximity_2005}.

Altermagnets have recently emerged as a class of symmetry-enforced compensated magnetic materials~\cite{smejkal_beyond_2022,bai_altermagnetism_2024}.  In parallel, fully compensated ferrimagnets (fFIMs)—a distinct class of filling-enforced compensated magnetic materials with vanishing net magnetization—have recently attracted growing interest for their electrically tunable spin splitting bands~\cite{liu_two-dimensional_2025,semboshiNewTypeHalfmetallic2022,kawamuraCompensatedFerrimagnetsColossal2024,yuan_nonrelativistic_2024,wurmehl_valence_2006,Hu_half,van_leuken_half-metallic_1995,MazinPRX2022,Guo2025LuttingerCB}, offering promising prospects for spintronic and topological applications. Notably, the coexistence of electrically tunable spin polarization and zero net magnetization not only meets the fundamental requirements for realizing topological superconductivity~\cite{qi_topological_2011, alicea_new_2012,stanescu_majorana_2013, sato_topological_2017}, but also enables all-electric control of topological phase transitions. These features establish fFIMs as a promising and scalable route toward magnetic-field-free topological quantum devices.
    \begin{figure}[t]
        \includegraphics[width=0.46\textwidth]{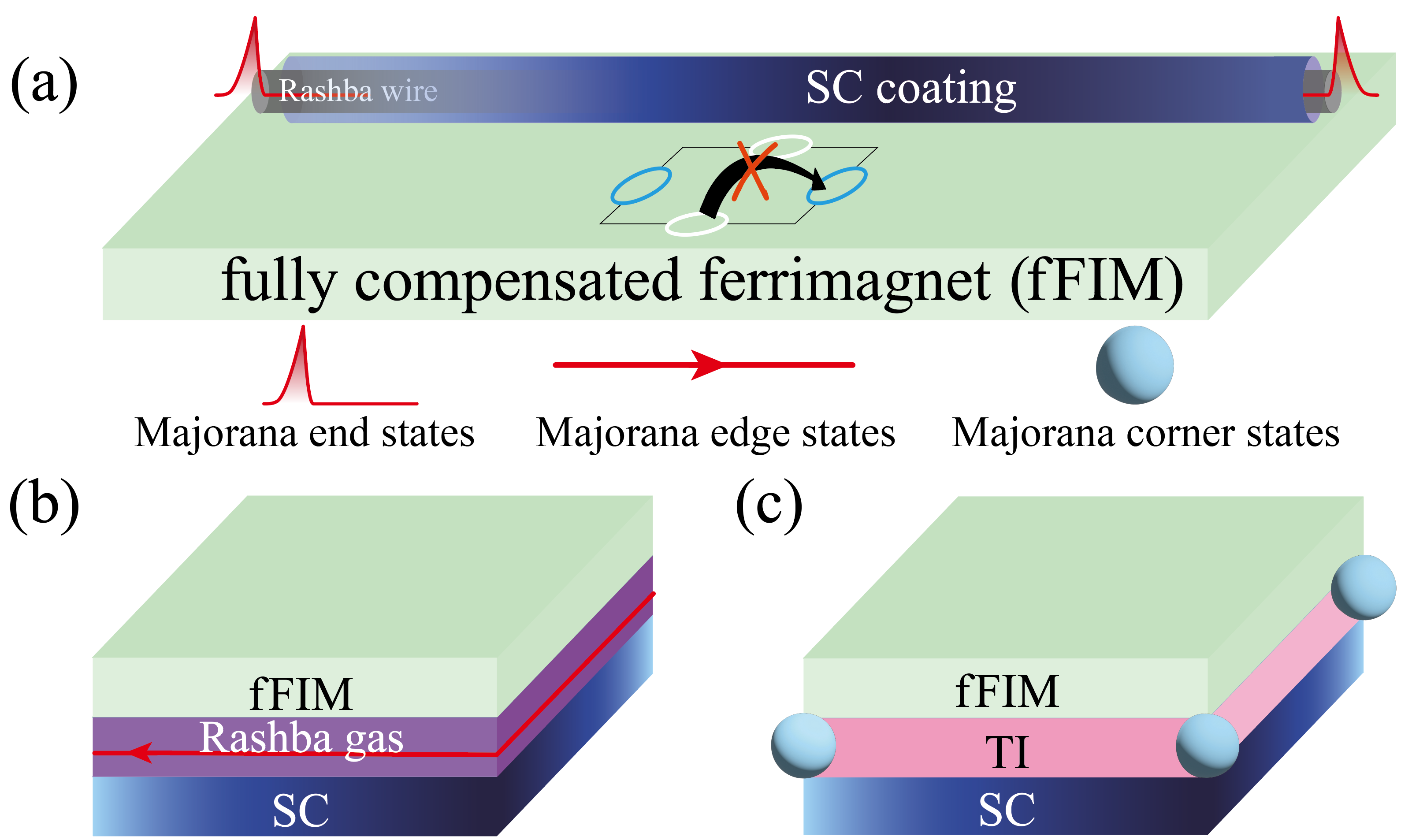}
        \caption{Proposal for tunable topological superconductivity by an intrinsic fFIM with electric field-controllable spin-band splitting.  (a) A 1D semiconductor nanowire with fFIM and a superconductor (SC) hosts  Majorana end states. (b) A  fFIM---Rashba electron gas---SC heterostructure hosts chiral Majorana edge states.  (c) Majorana corner modes emerge in a topological insulator (TI) sandwiched between a fFIM and a SC. }\label{fig:ill}
    \end{figure}

In this work, we demonstrate that heterostructures based on fFIMs provide a versatile platform for realizing tunable TSCs by both electric fields and the N\'eel vector, without net magnetization. This scheme eliminates the need for external magnetic fields or magnetic doping. By performing a detailed analysis, we identify three distinct TSC phases: (i) 1D TSCs hosting MZMs, (ii) 2D chiral TSCs with Majorana edge states (MESs), and (iii) higher-order TSCs featuring Majorana corner modes (MCMs). The system offers dual control, where electric fields universally induce topological phase transitions, while the orientation of the N\'eel vector governs the spatial distribution of MCMs in the higher-order TSCs and determines the chirality of MESs in the 2D conventional TSC phase. This combined control of electric fields and the N\'eel vector—achieved without introducing net magnetization—overcomes key challenges associated with conventional magnet-based approaches and provides a promising avenue for scalable Majorana devices.

\textit{Effective model of intrinsic fFIMs with controllable spin-band splitting.---} We first derive an effective model to describe the electric-field-tunable spin splitting in fFIMs. We adopt the lattice Hamiltonian with $d$-wave crystal potential, as taken from Ref.~\cite{liu_two-dimensional_2025,supp}, which is given by $H(\bfk)=2t_0(\cos k_x+\cos k_y)+V_1(\cos k_x-\cos k_y)\tau_z+J_0\bsf{s}\cdot\hat{\bsf{n}}\tau_z+V_0\tau_z-\mu$, where $t_0$ is the nearest-neighbor hopping, $\mu$ is the chemical potential, and $s_i$ ($\tau_j$) denote the Pauli matrices in spin (sublattice) space. The N\'{e}el vector, represented by a unit vector $\hat{\bsf{n}}=(\sin\theta\cos\varphi,\sin\theta\sin\varphi,\cos\theta)$, is tunable experimentally~\cite{meinert_electrical_2018,godinho_electrically_2018,bodnar_writing_2018,baltz_antiferromagnetic_2018,mahmood_voltage_2021,zhang_control_2022}. The second term represents a momentum-dependent crystal potential, the third term accounts for opposite magnetic moments on different sublattices, and the fourth term introduces an electric-field-controllable sublattice-staggered potential, which is critical for intrinsic fFIMs to break the equivalence of magnetic sublatices and not allowed for symmetry-enforced compensated magnets, such as  altermagnets~\cite{liu_two-dimensional_2025}.   A low-energy model is derived (see Supplemental Material (SM)~\cite{supp})
\begin{equation}
\begin{aligned}
\mathcal{H}^{\rm eff}(\bfk)&= 2t_0(\cos k_x + \cos k_y) - \mu\\
& + (V_0 + V_1(\cos k_x - \cos k_y))\bsf{s}\cdot\hat{\bsf{n}},
\end{aligned}
\end{equation}
which captures the electric-field-tunable spin-splitting of Fermi surface through the $V_0$.  The Fermi surface transitions from $\mathcal{C}_4$-symmetric $d$-wave  splitting at $V_0 = 0$ to $\mathcal{C}_4$-broken $s$-wave splitting at finite $V_0$~\cite{supp}. Thus, the third term in $\mathcal{H}^{\rm eff}(\bfk)$  faithfully describes fFIM band splitting, underpinning our subsequent analysis.

\begin{figure}[t]
    \includegraphics[width=0.48\textwidth]{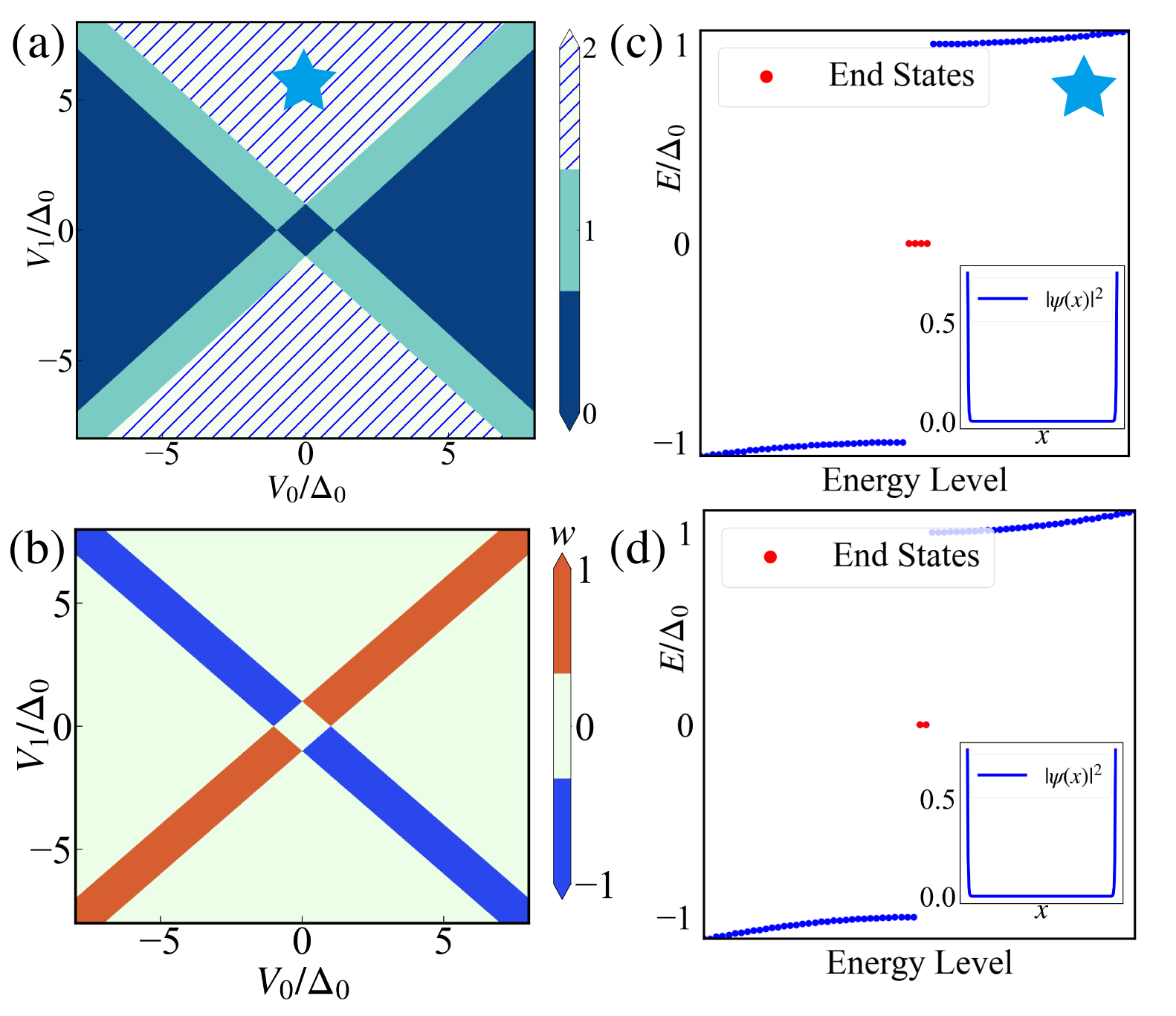}
    \caption{Topological phase diagram in 1D TSCs. (a) The topological invariant $\nu^++\nu^-$ calculated in the $(V_0, V_1)$ plane. (b) Winding number $w$ evolution in the same parameter space. (c) Spectrum with four zero modes; insets show wavefunctions localized at wire ends. (d) Spectrum with two zero modes; insets confirm end-state localization.  $\lambda_R=1$ is taken.  }\label{fig:1dnanowire}
\end{figure}

\textit{1D tunable TSCs with MZMs in nanowire.---}
We begin by considering a 1D semiconducting nanowire proximitized by an $s$-wave superconductor and placed on a fFIM substrate, as shown in Fig.~\ref{fig:ill}(a). To accurately describe the key features of this system, we employ a minimal Bogoliubov-de Gennes (BdG) Hamiltonian with the N\'eel vector aligned along the $z$-direction
\begin{equation}\label{eq:1dmodel}
\begin{aligned}
            H^{\rm 1D}_{\rm BdG}(k)=&\cs{V_0+V_1\cos k}s_z\gamma_z+\lambda_R\sin ks_y\gamma_z\\
            &+\Delta_0s_y\gamma_y-\mu\gamma_z,
\end{aligned}
\end{equation}
 where $\gamma_i$  are Pauli matrices in Nambu space. The first term originates from the fFIM substrate, where $V_0$ can be electrically tunable to modulate the band splitting from $d$-wave to $s$-wave~\cite{liu_two-dimensional_2025,supp}, $\lambda_R$ is the strength of the Rashba spin-orbit coupling (SOC). And $\Delta_0$ denotes the proximity-induced superconducting pairing.  The Hamiltonian $H^{\rm 1D}_{\rm BdG}(k)$ possesses particle-hole symmetry (PHS) $\mathcal{P}=\gamma_x\mathcal{K}$ (where $\mathcal{K}$ denotes complex conjugation) and, crucially, admits a purely real matrix representation~\cite{schnyder_classification_2008,tewari_topological_2012,tewari_topological_2012,supp}. The reality condition implies an effective time-reversal symmetry $ \widetilde{\mathcal{T}} = \mathcal{K} $ with $ \widetilde{\mathcal{T}}^2 = 1 $, combining with $ \mathcal{P} $ to generate a chiral operator $ \mathcal{C} = \gamma_x $. The system belongs to the BDI class, characterized by a $ \mathbb{Z} $ topological invariant, determined by the winding number $ w $~\cite{tewari_topological_2012}.   The topological phase diagram is systematically mapped by calculating the winding number $ w $ as a function of $V_0$ and $V_1$, as shown in Fig.~\ref{fig:1dnanowire}(b). A nonzero $w$ signals topologically nontrivial phases, characterized by zero-energy end states through bulk-boundary correspondence, as clearly demonstrated in the real-space spectrum of Fig.~\ref{fig:1dnanowire}(d). The corresponding wave functions, shown in the inset of Fig.~\ref{fig:1dnanowire}(d), exhibit exponential localization at the wire ends, providing direct evidence of MZMs in the TSC. Remarkably, the trivial-to-topological phase transition can be fully controlled via electrostatic gating of $V_0$, offering a highly tunable platform for realizing MZMs~\cite{supp}.

To investigate the electric-field tunability of topological states, we demonstrate that, at zero chemical potential ($\mu = 0$), our minimal model can be exactly mapped to the Kitaev chain~\cite{kitaev_unpaired_2001}. In this case, the Hamiltonian in Eq.~\eqref{eq:1dmodel} decomposes into two decoupled sectors due to the conserved quantity $\Pi = s_x\gamma_x$, which commutes with the Hamiltonian~\cite{supp}. Consequently, the Hamiltonian separates as $H = H^+ \oplus H^-$, where each subspace ($\Pi = \pm 1$) is described by an effective Hamiltonian
    \begin{equation}
        \widetilde{H}^\pm(k)=\cs{V_0+V_1\cos k\pm \Delta_0}\eta_z\mp\lambda_R \sin k\eta_y.
    \end{equation}
    Notably, each subspace Hamiltonian reduces to a 1D Kitaev chain, expressed in the Majorana basis as
    \begin{equation}
        \mathcal{A}^{\pm}(k)=\begin{pmatrix}
            0 & \zeta^{\pm}(k) \\
            -\zeta^{\pm*}(k) & 0
        \end{pmatrix},
    \end{equation}
    where $\zeta^{\pm}(k)=(V_0+V_1\cos k\pm\Delta_0)\mp i\lambda\sin k$. The condition $\det \mathcal{A}^{\pm}(k) = 0$ defines the topological phase boundary. The $\mathbb{Z}_2$ topological invariant is characterized by the Pfaffian as $(-1)^{\nu^\pm} = \text{sgn}[\text{Pf}(\mathcal{A}^\pm(0))] \cdot \text{sgn}[\text{Pf}(\mathcal{A}^\pm(\pi))]$, where $\nu^\pm = 0(1)$ indicates the topologically trivial (nontrivial) phase. By evaluating the $\mathbb{Z}_2$ invariants $\nu^\pm$ in the $(V_0, V_1)$ plane, we map out the composite phase diagram via $\nu_{\rm tot} = \nu^+ + \nu^-$ [Fig.~\ref{fig:1dnanowire}(a)]. The region with $\nu_{\rm tot} = 2$ hosts a quartet of zero-energy states with exponentially localized wavefunctions [Fig.~\ref{fig:1dnanowire}(c)]. Introducing a finite chemical potential induces hybridization among two MZMs at the same end, leading to an energy splitting $\delta E$, that signals the transition to a trivial phase~\cite{supp}.  This conclusion is further supported by the winding number calculation for $\mu\neq 0$~[Fig.~\ref{fig:1dnanowire}(b)], which rules out the existence of zero modes in the $\nu_{\rm tot}=2$ regime.

    Remarkably, the topological phase transition is not limited to electric-field manipulation; it can also be driven by reorienting the N\'eel vector, which significantly modifies the symmetry and band topology~\cite{li_majorana_2023,li_creation_2024}.  We find that electrically tunable MZMs persist when the N\'eel vector is aligned along the $x$-axis, while alignment along the $y$-axis drives the system into a topologically trivial phase~\cite{supp}. This dual tunability—enabled by both electric fields and N\'eel vector orientation—is conducive to engineering and controlling TSCs. Comprehensive analyses based on the full fFIM Hamiltonian $H(\mathbf{k})$, presented in the SM~\cite{supp}, further corroborate these findings.

\textit{2D tunable TSCs with chiral Majorana edge states---}
    We proceed to investigate a 2D heterostructure comprising a fFIM layer, a Rashba electron gas, and an $s$-wave superconductor, which supports robust chiral MESs along its boundary, as shown in Fig.~\ref{fig:ill}(b). We begin by considering the N\'eel vector along the $z$-direction ($\theta=0$) and construct a minimal model to describe the 2D TSC.
    \begin{equation}
        \begin{aligned}
            &H^{\rm 2D}_{\rm BdG}(\bfk)=\cs{V_0+V_1(\cos k_x-\cos k_y)}s_z\gamma_z\\
            &\qquad +\lambda_R\cs{\sin k_xs_y\gamma_z-\sin k_ys_x}+\Delta_0s_y\gamma_y-\mu\gamma_z,
        \end{aligned}
    \end{equation}
where the first term originates from the fFIM layer, the second term arises from the Rashba SOC in the electron gas, and $\Delta_0$ emerges from proximity-induced superconductivity.

    \begin{figure}[t]
    \includegraphics[width=0.48\textwidth]{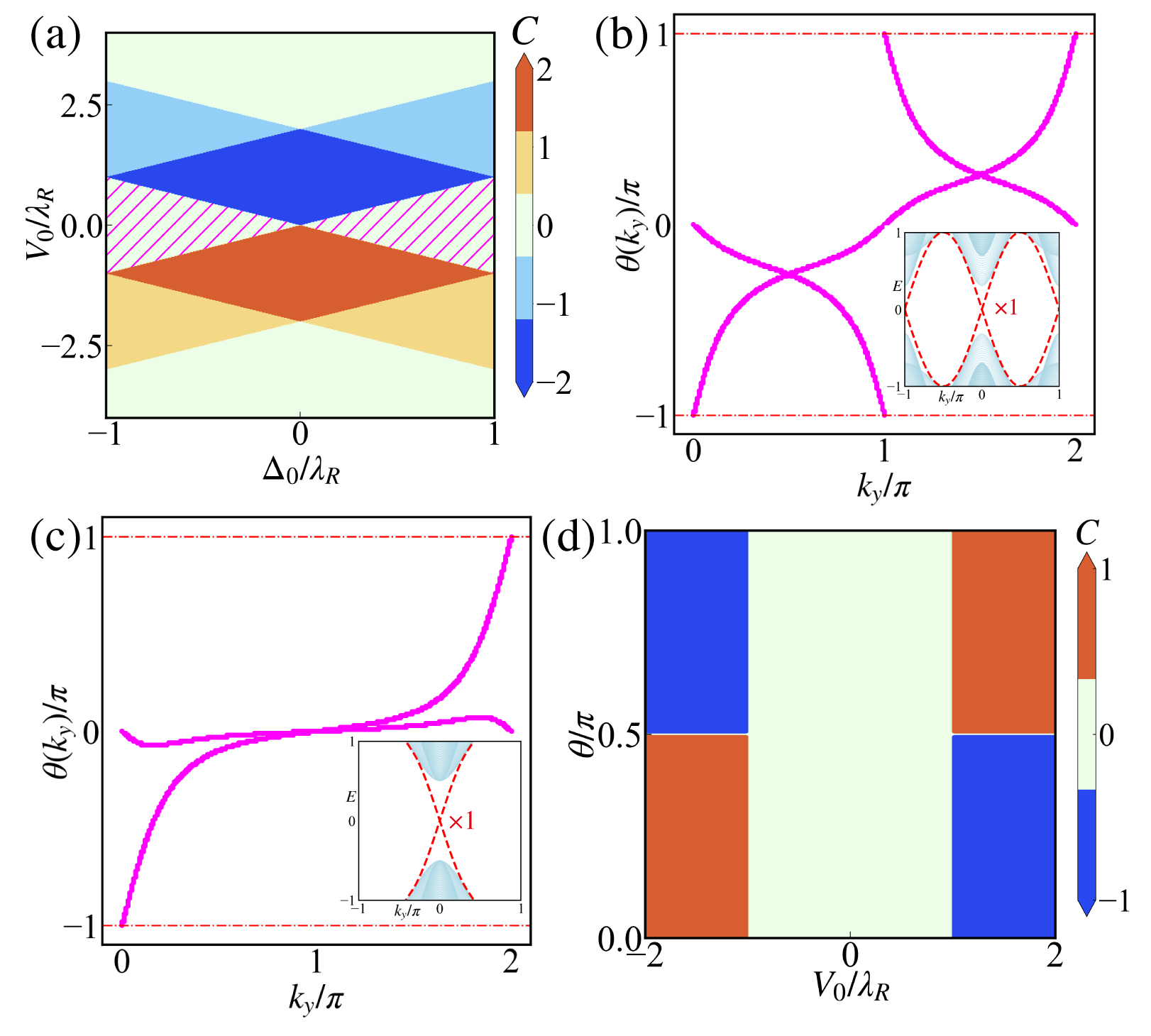}
    \caption{Topological phase diagram in 2D TSCs.  (a) The Chern number $(C)$ evolution in the  $(\Delta_0, V_0)$ plane. (b) Wilson loop spectrum calculated for the shaded region in (a), with the inset showing the edge states on a cylinder geometry. (c) Wilson loop is plotted for regions with Chern numbers $C = 1$, with corresponding edge states shown in the inset. (d) The tunable chirality of 2D TSC by the N\'eel vector and electric field in the fFIM. }\label{fig:2drashab}
\end{figure}
The system with PHS $\mathcal{P}=\tau_x\mathcal{K}$ belongs to class D, characterized by a $\mathbb{Z}$-valued Chern number $C$, which governs the emergence of chiral MEMs~\cite{schnyder_classification_2008}.    To clarify the topological nature of the system, we first consider $\mu = 0$. The calculated Chern number $C$ in the $(V_0, \Delta_0)$ plane yields a phase diagram with quantized values $C \in [-2, 2]$, as shown in Fig.~\ref{fig:2drashab}(a). In cylindrical geometry (open in $x$, periodic in $y$), the edge-state spectrum directly corresponds to the Chern number $C$. For $C = 1$, non-degenerate MEMs appear at $k_y = 0$ [Fig.~\ref{fig:2drashab}(c) inset], while for $C = 2$, doubly degenerate MEMs emerge at $k_y = 0$~\cite{supp}. Bulk-boundary correspondence is further confirmed through the Wilson loop calculation [Fig.~\ref{fig:2drashab}(c)], reinforcing the topological nature of the edge states.  Intriguingly,  in the shaded region where the Chern number vanishes~[Fig.~\ref{fig:2drashab}(a)], gapless edge modes persist at both $k_y = 0$ and $k_y = \pi$~[Fig.~\ref{fig:2drashab}(b) inset]. This phenomenon is explained by the Wilson loop, which reveals opposite nontrivial windings for the occupied bands, accounting for the presence of two counter-propagating chiral modes despite the net Chern number being zero. To further clarify the presence of MEMs in the shaded region despite the vanishing Chern number, we invoke a symmetry-based decomposition. Due to the conserved quantity $\Pi = s_x \gamma_x$, the Hamiltonian can be block-diagonalized into two decoupled sectors, each with distinct topological properties~[details in SM~\cite{supp}]. Each sector corresponds to a chiral $p \pm ip$-wave superconductor, with opposite signs for the pairing amplitude $\Delta_0$. The effective model shows that the Chern numbers for each sector are equal in magnitude but opposite in sign, $C^+ = -C^-$, resulting in a net $C = 0$. With a finite chemical potential ($\mu \neq 0$), the block-diagonal structure is lifted, leading to inter-sector hybridization and a slight modification of the phase diagram. MEMs at the same momentum exhibit a strong level repulsion and energy splitting, while at different momenta MESs retain their topological protection~\cite{supp}.

Building on previous analyses of topological phase control in 1D systems via the N\'eel vector, we extend this approach to manipulate the MEMs in 2D systems. We focus on the $ C = 1$ phase with a single chiral MEM. To explore the phase diagram, we compute the Chern number as a function of the polar angle of the N\'eel vector $(\theta)$ and electric field $(V_0)$ in the fFIM, as shown in Fig.~\ref{fig:2drashab}(d). When the N\'eel vector lies within the $x$-$y$ plane ($\theta = \pi/2$), the system remains topologically trivial. Remarkably, the introduction of an out-of-plane component ($\theta \neq 0$) drives the system into a topological phase hosting chiral MEMs. The chirality of these edge states is tunable by both the polar angle $\theta$ and the gate voltage $V_0$.  This tunability provides a controllable mechanism to manipulate the chirality of edge modes, which is highly relevant for potential quantum computing applications~\cite{alicea_new_2012}. These results, obtained from the minimal model, are consistent with those from the full model, as detailed in the SM~\cite{supp}.

\begin{figure}[t]
        \includegraphics[width=0.5\textwidth]{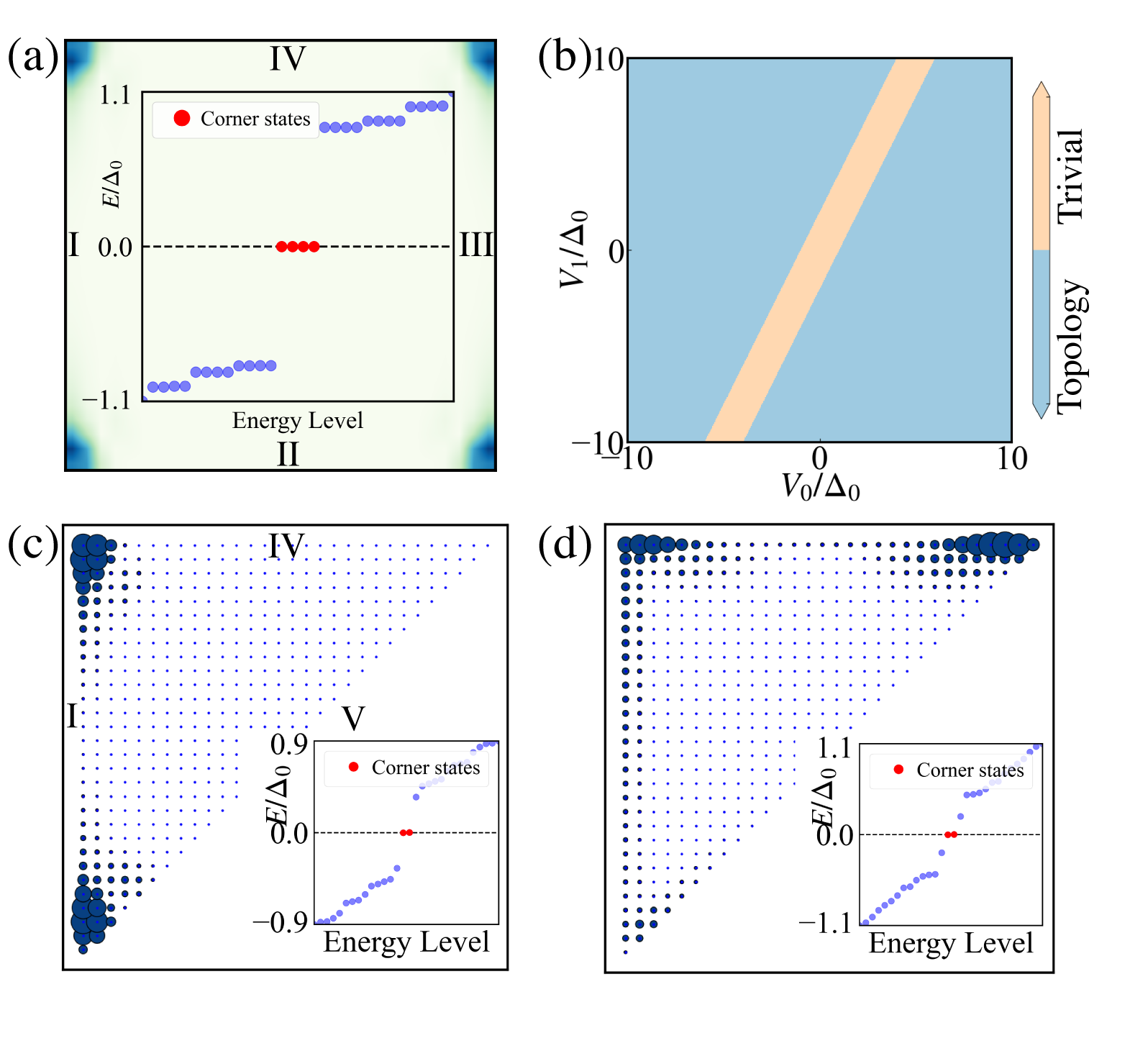}
        \caption{Phase diagram and MCMs evolution in the higher-order TSC.  (a) Spatial localization of one MZM per corner, with the spectrum (inset) showing zero-energy eigenvalues.  (b) Phase diagram in the $(V_0, V_1)$ parameter space.  (c) Energy spectrum (inset) of an isosceles triangle structure with two MCMs, and their wavefunction density profile for the N\'eel vector along the $x$ direction. (d) Analogous to (c),  for the N\'eel vector along the $y$ direction. Parameters: $t_0=1.0,m_0=1.0,\lambda=1.0,V_0=0.5,V_1=0.5, \Delta_0=0.2.$ }\label{fig:hotsc}
    \end{figure}

\textit{Higher-order TSCs with tunable Majorana corner states.---} The N\'eel vector governs not only first-order topological states but also plays a crucial role in controlling higher-order topological phases~\cite{li_majorana_2023,li_creation_2024}. We construct a sandwich structure, as shown in Fig.~\ref{fig:ill}(c), and develop a minimal model
    \begin{equation}\label{eq:sc}
        \begin{aligned}
            H_{\rm BdG}(\bfk)&=M(\bfk)\sigma_z\gamma_z+\lambda\cs{\sin k_x s_y\sigma_x\gamma_z-\sin k_y s_x\sigma_x}\\
            &+\cs{V_0+V_1(\cos k_x-\cos k_y)}\bsf{s}\cdot\bsf{\hat{n}}\gamma_{z(0)}\\
            &+\Delta(\bfk)s_y\gamma_y.
        \end{aligned}
    \end{equation}
The first line represents the topological insulator with a nontrivial $\mathbb{Z}_2$ invariant, the second incorporates the fFIM, and the third accounts for the superconducting proximity effect.  For illustration, we consider $\mu = 0$ and the N\'eel vector along the $x$-axis ($\theta = \pi/2$, $\varphi = 0$). With pairing and the fFIM, the system transitions into a higher-order topological superconductors (HOTSC) phase, characterized by corner-localized MZMs, as shown in Fig.~\ref{fig:hotsc}(a).

The emergence of MCMs can be understood through edge theory. We  derived the low-energy Hamiltonian for all edges ($\ell$=I---IV), yielding the compact form~\cite{supp}
\begin{equation}\label{eq:edge_hamiltonian}
        H^{\rm edge}(\ell) = -iA(\ell)\eta_z\kappa_0\partial_\ell - M(\ell)\eta_y\kappa_y + \mathcal{V}(\ell)\eta_y\kappa_0.
\end{equation}
Using the quantity $\kappa_y = \pm 1$, the edge Hamiltonian $H^{\rm edge}(\ell)$ can be block-diagonalized as $H = H^+ \oplus H^-$, with each subspace governed by an effective Dirac Hamiltonian:
$ H^\pm = iA(\ell)\eta_z\partial_\ell + \tilde{M}^\pm(\ell)\eta_y. $
Here, the Dirac mass $\tilde{M}^\pm(\ell)$ on edges I--–IV takes the form
$ \tilde{M}^\pm = \{\mp\Delta_0, \ \mp\Delta_0 \mp (V_0 - V_1|m|/t_0), \ \pm\Delta_0, \ \pm\Delta_0 \pm (V_0 - V_1|m|/t_0)\}. $   MCMs emerge at the domain walls where the Dirac masses change sign across adjacent edges in either $H^+$ or $H^-$ subspace. In the case of Fig.~\ref{fig:hotsc}(a), an $x$-oriented N\'eel vector induces mass terms on edges II and IV, producing sign reversals in both $H^+$ and $H^-$ subspaces. The resulting mass domain walls, appearing at distinct corners in the two sectors, each binds a localized MZM~\cite{supp}.  Based on our edge theory, we analyze the dependence of the Dirac mass sign on the parameters $(V_0, V_1)$ and present the phase diagram shown in Fig.~\ref{fig:hotsc}(b). Our results demonstrate that electric fields can effectively control the generation and annihilation of MCMs.

To explore the tunability of MCMs by the N\'eel vector, we consider an isosceles right triangular geometry. Figures~\ref{fig:hotsc}(c,d) illustrate the spatial redistribution of MCMs under two orthogonal N\'eel vector configurations with azimuthal angles $\varphi = 0$ and $\varphi = \pi/2$. In Fig.~\ref{fig:hotsc}(c), for $\varphi = 0$, the MCMs are localized at the terminations of edge I. In contrast, in Fig.~\ref{fig:hotsc}(d), for $\varphi = \pi/2$, the MCMs shift to the terminations of edge IV.  This controllable switching of MCM positions demonstrates the N\'eel vector as an effective means for manipulating MCMs. The observed tunability of MCMs can be naturally understood through edge theory.  For the triangular structure, the edge Hamiltonian in each sector is given by $H^\pm = iA(\ell)\eta_z\partial_\ell + \tilde{M}^\pm(\ell)\eta_y$, with edge-dependent Dirac masses $\tilde{M}^\pm(\ell = \text{I, IV, V}) = \{\pm\Delta_0 \pm V_0 + V_1|m|/t_0,\ \pm\Delta_0 \mp V_0 + V_1|m|/t_0,\ \pm\Delta_0 \pm V_0\}$. When the N\'eel vector aligns along the $x$-axis ($\varphi = 0$), mass sign reversals occur at corners I–IV and I–V, creating domain walls that bind MCMs. Rotating the N\'eel vector to the $y$-axis ($\varphi = \pi/2$) removes the I–V domain wall and generates a new one at corner IV–V, resulting in the movement of MCMs.

In addition to the electric field and N\'eel vector tuning, a nonzero chemical potential also influences the formation of MCMs by coupling the two subspaces $H^\pm$. While MCMs remain robust when MZMs from $H^+$ and $H^-$ appear at different corners, hybridization occurs when they merge at the same corner, leading to energy splitting, similar to the level repulsion observed in the 1D case~\cite{supp}.

\textit{Conclusion and discussion.---} In summary, we demonstrate that fFIM-based heterostructures provide a tunable platform for  TSC, offering both electrical and N\'eel vector control. The electric field serves as the primary driver for topological phase transitions, enabling band splitting within the fFIM layer without the need for external magnetic fields. Additionally, the N\'eel vector orientation plays a key role, enabling (i) deterministic control over the chirality of MEMs in 2D TSC and (ii) spatial manipulation of MCMs in HOTSC phase. This dual control mechanism—where the electric field governs bulk topological properties and the N\'eel vector modulates boundary states—addresses the challenge of balancing tunability with superconducting integrity in magnetized systems. The ability to operate without external magnetic fields and reconfigure the system via voltage makes fFIM-based heterostructures promising candidates for non-Abelian quantum devices.  Inspired by the electric-field-tunable spin splitting in fFIMs, our proposal, while focused on fFIMs with a $d$-wave crystal potential, is equally applicable to fFIMs with other momentum-dependent crystal potentials, such as $g$-wave and $i$-wave~\cite{li_creation_2024,liu_two-dimensional_2025}.

Experimental detection of Majorana modes can be achieved through dimension-specific probes tailored to the distinct topological phases. In both 1D TSC and 2D HOTSC, scanning tunneling microscopy is expected to reveal quantized zero-bias conductance peaks at $2e^2/h$, arising from resonant Andreev reflection at localized MZMs~\cite{law_majorana_2009, wimmer_quantum_2011, das_zero-bias_2012, jack_observation_2019}. In 2D chiral TSC phases, MESs can be identified through thermal transport measurements exhibiting half-quantized thermal Hall conductance~\cite{jezouin_quantum_2013, banerjee_observation_2018}, or by observing the $4\pi$-periodic Josephson effect in phase-biased junctions~\cite{fu_josephson_2009, fornieri_evidence_2019}. These complementary spectroscopic and transport signatures would provide evidence for the presence of Majorana modes.

\textit{\color{red}Acknowledgments.---}
The authors thank Jinyi Duan, Ling Bai, and Chuanchang Zeng for helpful discussions. The work is supported by the  Science Fund for Creative Research Groups of NSFC (Grant No. 12321004), the NSF of China (Grant No. 12374055), and the National Key R\&D Program of China (Grant No. 2020YFA0308800).

    \bibliography{ref}
\end{document}